\documentclass{article}

\usepackage[T1]{fontenc}
\usepackage[utf8]{inputenc}
\usepackage[margin=1in,top=1in,bottom=3cm]{geometry}
\usepackage{perpage}
\MakePerPage{footnote}
\usepackage{epsfig}
\usepackage{mathtools}
\usepackage{latexsym}
\usepackage{amsmath}
\usepackage{mathrsfs}
\usepackage{yfonts}
\usepackage{makeidx}
\usepackage{graphicx}
\usepackage{slashed}
\usepackage{hyperref}
\usepackage[titletoc]{appendix}

\usepackage{chngcntr}
\usepackage{bm}
\usepackage{dsfont}
\usepackage{multicol}
\usepackage{nameref}
\usepackage{textcomp}
\usepackage{changepage}
\usepackage{lettrine}
\usepackage{enumitem}
\usepackage[sortcites,backend=bibtex,style=numeric,sorting=anyvt,doi=false,url=false,giveninits=true,isbn=false]{biblatex}
\AtEveryBibitem{\clearfield{month}}
\AtEveryBibitem{\clearfield{day}}
\usepackage{tikzsymbols}
\hypersetup{citecolor=blue,colorlinks=true,linkcolor=blue,pdfstartview={FitH},linktoc=page,pdftitle={Maxwell-Decay-on-DSRN-BH.pdf}pdfauthor={Mokdad Mokdad}}
\usepackage{amsthm,amssymb}
\usepackage{subcaption}
\usepackage{mathtools}
\mathtoolsset{showonlyrefs=true}

\newtheorem{thm1}{Theorem}%[section]
\newtheorem{prop1}[thm1]{Proposition}

\newtheorem{note}[thm1]{Remark}

 \newtheoremstyle{TheoremNum}
        {\topsep}{\topsep}              %%% space between body and thm
        {\itshape}                      %%% Thm body font
        {}                              %%% Indent amount (empty = no indent)
        {\bfseries}                     %%% Thm head font
        {.}                             %%% Punctuation after thm head
        { }                             %%% Space after thm head
        {\thmname{#1}\thmnote{ \bfseries #3}}%%% Thm head spec
    \theoremstyle{TheoremNum}

     \newtheoremstyle{TheoremNum}
        {\topsep}{\topsep}              %%% space between body and thm
        {\itshape}                      %%% Thm body font
        {}                              %%% Indent amount (empty = no indent)
        {\bfseries}                     %%% Thm head font
        {.}                             %%% Punctuation after thm head
        { }                             %%% Space after thm head
        {\thmname{#1}\thmnote{ \bfseries #3}\thmnumber{}}%%% Thm head spec
    \theoremstyle{TheoremNum}

\renewcommand{\d}{\mathrm{d}}

\newcommand{\R}{\mathbb{R}}

\newcommand{\Sph}{{\mathcal{S}^2}}
\newcommand{\Sphw}{{\mathcal{S}^2_{\omega}}}
\newcommand{\dl}{\partial}

\newcommand{\hf}{\frac{1}{2}}

\newcommand{\hook}{{\setlength{\unitlength}{11pt}   % adjust pt size here
                   \begin{picture}(.833,.8)
                   \put(.15,.08){\line(1,0){.35}}
                   \put(.5,.08){\line(0,1){.5}}
                   \end{picture}}}
\newcommand{\scrh}{{\mathscr H}}
\newcommand{\scrs}{{\mathscr S}}
\newcommand{\hlm}{{\scrh}^L_{r_{_-}}}
\newcommand{\hrm}{{\scrh}^R_{r_{_-}}}
\newcommand{\hlp}{{\scrh}^L_{r_{_+}}}
\newcommand{\hrp}{{\scrh}^R_{r_{_+}}}              
\newcommand{\hpz}{\hat{\phi}_0}
\newcommand{\hpo}{\hat{\phi}_1}
\newcommand{\hl}{\hat{l}}
\newcommand{\hn}{\hat{n}}

\renewcommand{\S}{\mathbb{S}}
               
\makeindex

\title{Conformal Scattering and the Goursat Problem for Dirac Fields in the Interior of Charged Spherically Symmetric Black Holes}
\author{Mokdad Mokdad\thanks{Université Grenoble Alpes -- Institut Fourier -- 100 rue des maths, 38610 Gières, France. \newline email: mokdad.al.mokdad@gmail.com - mokdad.mokdad@univ-grenoble-alpes.fr }}

\begin{document}

\maketitle

\begin{abstract}
We construct a conformal scattering theory for Dirac fields in the interior of a Reissner-Nordström-like black hole, between the black hole event horizon and the Cauchy horizon. The main result is a resolution of the  characteristic Cauchy problem for the Dirac equation on the horizons by solving a system of wave equations and re-interpreting its solution as the components of the required Dirac solution.
\end{abstract}

{\bf Keywords.} Conformal Scattering theory, Black hole interior, Reissner-Nordström metric, Goursat Problem, Dirac equation.

\vspace{0.1in}

{\bf Mathematics subject classification.} 35Q41, 35Q75, 83C57.

\tableofcontents

\section{Introduction}

Recently, the author and his collaborators \cite{hafner_scattering_2020} obtained a complete scattering theory for Dirac fields in the interior of Reissner-Nordström-like black holes, using an analytic framework which was then re-interpreted geometrically. The propose of this short subsequent note is to show that the geometric results in the same settings can be directly obtained using what we will refer to as the method of ``waves re-interpretation'' for solving the characteristic Cauchy problem (also known as the Goursat problem), and to show that the associated scattering theory is immediate from it. We follow the tradition of calling the scattering theory we construct here a << conformal scattering>> theory even though no conformal rescaling is needed, as we still apply the rest of the geometric construction in conformal scattering, namely, the usage of geometric estimates and the direct resolution of the Goursat problem. In addition, situations requiring conformal rescaling  may very well be amenable to such a construction.  

The importance of studying the scattering of fields in general relativity is now an established fact. In particular, studying scattering in the interior regions is expected to be important for the Cauchy horizon instability problem of Roger Penrose cosmic censorship conjecture, thought what is known as the gravitational blue shift at the horizon.   The literature on scattering theory is very vast and reviewing it is out of the scope of this work. The reader can find references for literature reviews in \cite{hafner_scattering_2020}, and an overview of the history of conformal scattering in \cite{mokdad_conformal_2019}. Howbeit, there are few scattering results in the interior regions of black holes. Up to our knowledge, the scattering theory for waves obtained by C. Kehle and Y. Shlapentokh-Rothman \cite{kehle_scattering_2019}, and the aforementioned results by D. Häfner, J.-P. Nicolas and the author \cite{hafner_scattering_2020}, are the only complete scattering theories in the interiors of black holes.    

The method of  ``waves re-interpretation'' was first used by the author in \cite{mokdad_conformal_2019} to solve the Goursat problem for Maxwell fields in the exterior region, and then by T.X. Pham \cite{pham_peeling_2017} for a general spin-$\frac{n}{2}$ fields on Minkowski spacetime. The idea of this method is a successive applications of the operator in study to alternate between a system of wave equations and the original equations (possibly with perturbations). The transfer from the Dirac (Maxwell or spin-$\frac{n}{2}$) Goursat problem to a wave Goursat problem is done by applying a Dirac operator a second time to the Dirac equations to obtain a system of wave equations. We then use the already established theory for the well-posedness of the latter, including regular perturbations up to the first order. In particular, the study in \cite{hormander_remark_1990} is sufficient for our purposes. The second step is to reinterpret the wave solution as a Dirac field, which again is done by applying the Dirac operator once more and using the well-posedness of the wave Goursat problem. In  \cite{mokdad_conformal_2019}, the singularity of $i^+$ was in the future of the initial Cauchy hypersurface of $t=0$, where the  initial data is taken. In that case, involved decay estimates \cite{mokdad_decay_2020}  for Maxwell fields were needed first before establishing the energy estimates. Thanks to the conserved current of the Dirac equation, the required estimates are immediate. However, in the interior region, the $i^+$ singularity is in the of past of the Goursat data, where the initial hypersurface ``meets'' the horizons. To deal with that, we use a technique similar to the one that was used by J.-P. Nicolas in  \cite{nicolas_conformal_2016} to deal with the $i_0$ singularity in the exterior of the Schwarzschild black hole.     

We expect this method to be robust and general enough to be applied in different situations. In particular we strongly expect that it is applicable in the asymmetric cases of Kerr black holes. Also, obtaining a full study for the Goursat problem of the general spin equation on a generic background may be plausible indeed. These prospects will be the focus of future projects.

The layout of the paper is as follows. First, in section \ref{Sect:Settings}, we present the geometric background describing the interior region of a Reissner-Nordström-like black hole between the black hole event horizon and the Cauchy horizon. We also gather in this section the elements we need for the Newman-Penrose formalism, namely a collection of tetrads adapted to our geometry and there associated spin-frames. Scattering for Dirac fields is discussed in section \ref{Sect:Scattering}. There, we introduce the Dirac equation we will be studying together with its current, and we state their properties that are required for us. The trace and scattering operators are defined in Proposition \ref{Prop:existenceoftraceops} and Theorem \ref{theoremScattering}. Section  \ref{Section:Gourast} contains the main result of the paper: the resolution of the Goursat problem for the Dirac equation on the horizons. We start this section by obtaining the system of wave equations mentioned above in the description of our method. Theorem \ref{theorem:Goursat} states the main result, and the waves re-interpretation is done in its proof.

\paragraph{Acknowledgement} The author would like to thank J.-P. Nicolas for a valuable discussion regarding the singularity at $i^+$.  

\paragraph{Notations and conventions} Since some of our conventions and of the preliminaries we need here are in common with the previous work in \cite{hafner_scattering_2020}, we will recall a minimum amount in section \ref{Sect:Settings}  and we refer the reader to \cite{hafner_scattering_2020} and the references within for more on them.  However, we will use a slightly different notation for partial derivatives. We will use the notation $\dl_y$ to mean $\frac{\dl}{\dl y}$ for the partial derivative with respect to a generic variable $y$.

\section{Settings}\label{Sect:Settings}
 
\subsection{The Geometric Framework}

The spacetimes we consider are Reissner-Nordström-(anti-)de-Sitter like black holes, and we work in the interior region situated between the two inner horizons --- the Cauchy and the event horizons. We model such a region by $\R_x\times]r_-,r_+[_r\times\mathcal{S}_{\omega}^2$ with $0<r_-<r_+<+\infty$, equipped with a metric of the form 
\begin{equation}\label{RNDSmetric}
\mathbf{g}=-f(r)\d x^2+\frac{1}{f(r)}\d r^2-r^2\d \omega^2 \; , \qquad   \mathrm{with} \quad \d \omega^2=\d \theta^2 + \sin(\theta)^2\d \varphi^2 \; ,
\end{equation}
and $f\in\mathcal{C}^\infty\left([r_-,r_+]\right)$ positive on $]r_-,r_+[$, and $r_\pm$ are its only zeros in $[r_-,r_+]$. We use prime to indicate derivation with respect to $r$, such as $f':=\dl_r f$.

In terms of a Regge-Wheeler coordinate $t$, defined by $\frac{\d r}{\d t}=-f$ and an arbitrary origin point for $t=0$, the metric becomes 
\begin{equation}\label{RNDSmetricINRstar}
\mathbf{g}=f(r)(\d t^2 - \d x^2)-r^2\d \omega^2, 
\end{equation}
and it is now define on $\mathcal{M}=\R_x\times\R_t\times\mathcal{S}_{\omega}^2$. We choose to orient $\mathcal{M}$ so that $(t,x,\theta,\varphi)$ is a positively oriented coordinate chart, and we fix a time orientation on $\mathcal{M}$ by declaring the timelike vector field $\dl_t$ to be future oriented.  Note that $r$ is a smooth, strictly decreasing function of $t$ on $\R$ and $r\rightarrow r_\mp$ as $t\rightarrow \pm \infty$. We denote by $\Sigma_t$ the level hypersurfaces of the time function $t$, and we denote its future-oriented unit normal by $\eta_t =f^{-\hf}\dl_t$.

\begin{figure} 
	\centering
	\includegraphics[scale=1.2]{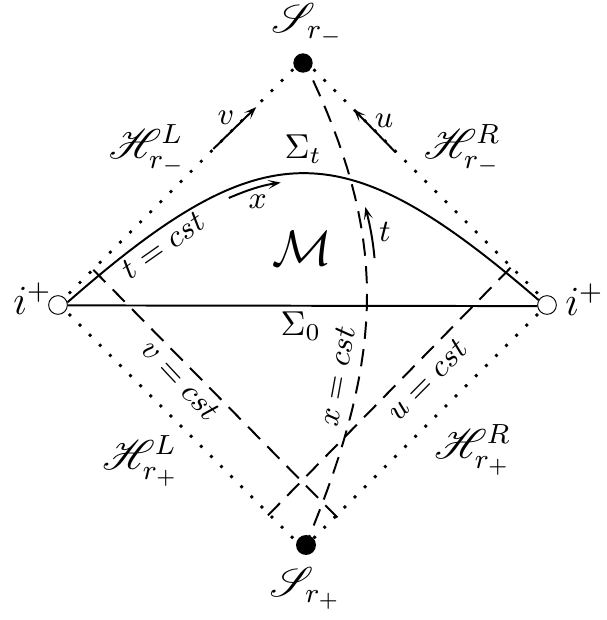}
	\caption{\emph{A Penrose-Carter conformal diagram  of $\mathcal{M}.$ 
			%and its boundary with the different coordinates on them. The small arrows indicate the positive direction of each variable.
		}}
	\label{fig:diagM}
\end{figure}

The pair $(\mathcal{M},\mathbf{g})$ can be extended to cover the values $r=r_\pm$ by introducing the Eddington-Finkelstein variables $u = t-x$ and $v = t+x$. In terms of the coordinate systems $(u,r,\omega)$ and $(v,r,\omega)$, the metric has the same expression, namely:
\begin{align}
g &= - f(r) \d u^2 - 2 \d u \d r - r^2 \d \omega^2 \, , \label{REFMet}\\
&= - f(r) \d v^2 - 2 \d v \d r - r^2 \d \omega^2 \, , \label{AEFMet}
\end{align} 
%
%Another useful form of the metric is in the coordinate system $(u,v,\omega )$
%%
%\begin{equation} \label{uvMet}
%g = f(r) \d u \d v - r^2 \d \omega^2 \, .
%\end{equation}
showing that $\mathbf{g}$ extends as a smooth non-degenerate metric to $\mathcal{M}_u:=\R_u\times [r_-,r_+]_r  \times \mathcal{S}_{\omega}^2$ and to  $\mathcal{M}_v:=\R_v\times [r_-,r_+]_r  \times \mathcal{S}_{\omega}^2$. Thus, four smooth null hypersurfaces can be added to $\mathcal{M}$ in the regions $\{ r=r_\pm\}$, referred to respectively as the left and right inner and outer horizons:

\begin{align*}
	\hlm &:= \R_v \times\{ r=r_-\}  \times \mathcal{S}_{\omega}^2 \, ,\\
	\hrm &:= \R_u \times\{ r=r_-\} \times  \mathcal{S}_{\omega}^2 \, ,\\
	\hlp &:=  \R_u \times\{ r=r_+\} \times \mathcal{S}_{\omega}^2 \, ,\\
	\hrp &:= \R_v \times \{ r=r_+\} \times \mathcal{S}_{\omega}^2 \, .
\end{align*}
We refer to $\scrh_{r_-}=\hlm \cup \hrm$ as the Cauchy (or inner/future) horizon and to $\scrh_{r_+}=\hlp \cup \hrp$  as the event (or outer/past) horizon. The spacetime can be further extended so that the left and right components of each horizon meet in a smooth $2$-sphere, called the bifurcation sphere, which we denote by $\scrs_{r_+}$ for the outer horizon and $\scrs_{r_-}$ for the inner horizon. These can be constructed by the usual procedure using Kruskal-Szekeres-type coordinates. As we will not be doing explicit calculations on these 2-spheres, we do not give the precise expressions of such coordinates\footnote{See for example \cite{hafner_scattering_2020,mokdad_reissnernordstromsitter_2017} for the expressions of the coordinates on the bifurcation spheres.}, and we are content with noting that, say $\scrs_{r_-}$,  is reached as $v \rightarrow +\infty$ along $\hlm$, as $u \rightarrow +\infty$ along $\hrm$ or going towards the future (i.e. $t\rightarrow +\infty$) along all lines of fixed $x$ and $\omega$. See Figure \ref{fig:diagM} for a Penrose diagram of the spacetime $\mathcal{M}$ and its boundaries.
Thus, the future boundary of $\mathcal{M}$ is the set
$ \{ r=r_- \} = \scrh_{r_-} \cup \scrs_{r_-}$
and its past boundary is
$ \{ r=r_+ \} =\scrh_{r_+}  \cup \scrs_{r_+}$, and we denote by $\bar{\mathcal{M}}$ the spacetime $\mathcal{M}$ with these two boundaries attached to it. 

The regions reached as $x\rightarrow\pm \infty$ along a curve of fixed $t$ (the regions labelled $i^+$ where the future and past horizons ``meet'' in Figure \ref{fig:diagM}) are not compactified, and are regarded as asymptotic regions for our extend spacetime. The $i^+$ corresponds to future timelike infinity in a sub-extremal Reissner-Nordström-(anti-)de-Sitter black hole when two distinct inner horizons exist. In the case of a Reissner-Nordström black hole, the cosmological constant is zero, $\Lambda=0$, and the condition for having two horizons is simply the mass being greater than the absolute value of a non-zero charge. 
% In all three cases, the metric is given on $\R_{\tilde{t}} \times ]0,+\infty [_r \times \Sph^2_\omega$ by
%
%\begin{equation} \label{DSRNMet}
%g = F \d \tilt^2 - \frac{1}{F} \d r^2 - r^2 \d \omega^2 \, ,~ F = 1 - \frac{2M}{r} + \frac{Q^2}{r^2} - \Lambda r^2 \, ,
%\end{equation}
%
%where $M$ and $Q$ are the mass and charge of the black hole and $\Lambda \in \R$ is the cosmological constant. We place ourselves between two consecutive simple horizons, corresponding to $0<r_-<r_+$, such that their respective surface gravities satisfy $\kappa_- < 0 < \kappa_+$. For $\Lambda = 0$ (Reissner-Nordström), this requires $0<\vert Q \vert <M$ and we have $r_\pm = M \pm \sqrt{M^2-Q^2}$. 
For $\Lambda >0$ (Reissner-Nordström-de Sitter), the required conditions are detailed in \cite{mokdad_reissnernordstromsitter_2017}, and for $\Lambda <0$ (Reissner-Nordström-anti-de Sitter) the conditions are given in \cite{hafner_scattering_2020}. Note that in all of these three cases, the charge is non-zero when two inner horizons exist.

%The non-zero Christoffel symbols are:
%\begin{enumerate}[label={\alph*)}]
%	\item in the $(x,r,\theta,\varphi)=(x^0,x^1,x^2,x^3)$:
%\begin{gather}
%\Gamma^0_{01}=-\Gamma^1_{11}=\frac{f'}{2f} ~~~ ;~~~ \Gamma^1_{00}=\frac{ff'}{2} ~~~ ; ~~~ \Gamma^1_{22}=rf ~~~ ; ~~~ \Gamma^1_{33}=rf\sin(\theta)^2  \nonumber \\ \Gamma^2_{12}=\Gamma^3_{13}=\frac{1}{r} ~~~ ; ~~~ \Gamma^2_{33}=-\cos(\theta)\sin(\theta) ~~~;~~~ \Gamma^3_{23}= \cot(\theta) ~~ . \label{Crstflsymb}
%\end{gather}
%
%\item in the $(x,t,\theta,\varphi)=(\tilde{x}^0,\tilde{x}^1,\tilde{x}^2,\tilde{x}^3)$: 
%\begin{gather}
%\tilde{\Gamma}^0_{01}=\tilde{\Gamma}^1_{00}=\tilde{\Gamma}^1_{11}=-\frac{f'}{2}~~~ ;~~~ \tilde{\Gamma}^1_{22}=-r ~~~ ;~~~\tilde{\Gamma}^1_{33}=-r\sin(\theta)^2  \nonumber       \\
%\tilde{\Gamma}^3_{13}=\tilde{\Gamma}^2_{12}=-\frac{f}{r}~~~ ;~~~\tilde{\Gamma}^2_{33}=-\cos(\theta)\sin(\theta)~~~ ;~~~\tilde{\Gamma}^3_{23}=\cot(\theta).\label{Crstflsymbinr*}
%\end{gather}
%\end{enumerate}

\subsection{Spinors and Newman-Penrose Tetrads}

In many occasions throughout the paper, we will use the abstract index notation as well as the Newman-Penrose formalism. More on these subjects can be found in   \cite{penrose_spinors_1987} . For example, a vector field $V$ will also have an indexed version $V^a$, and $V_a$ will be the 1-form obtain by lowering the index using the metric. A Newman-Penrose tetrad on a spacetime is a set of four null vector fields $\{ l,n,m,\bar{m}\}$, two of which are real and the complex vectors are conjugate of one another, that form a local basis of the complexified tangent bundle of the space-time manifold.   It is said to be normalized if $ l_an^a = -m_a\bar{m}^a = 1$. 

If we assume that the spacetime is globally hyperbolic, then it admits a spin-structure, and we denote by $\S^A$  the bundle of spin-vectors over it, and by $\S_A$ its dual bundle. We also denote by $\S^{A'}$ and $\S_{A'}$ their respective complex conjugate bundles. A spin-frame is a local basis $\{ o^A , \iota^A \}$ of $\S^A$ normalized by the condition $o_A \iota^A =1$. Note that because of the anti-symmetry of the symplectic structure on $\S^A$, $\iota_Ao^A=-1$. A spinor  field $\phi_A$ (or $\phi^A$) is a section of the bundle $\S_A$ ( respectively $\S^A$), and one can write its components in the spin-frame $\{ o , \iota \}$ as

\begin{equation}\label{componentsintetrad}
\phi_0 =-\phi^1 = \phi_A o^A \quad\textrm{and} \quad \phi_1 = \phi^0= \phi_A \iota^A \, , 
\end{equation} 
and so we have, for example,
\begin{equation}\label{componentdecompofspinor}
	\phi_A=\phi_1 o_A -\phi_0\iota_A \, . 
\end{equation}
The components $\phi_{0'}$ and $\phi_{1'}$ of $\bar{\phi}_{A'}$ are defined similarly, and are in fact the complex conjugate of $\phi_0$ and $\phi_1$ respectively. 

To a normalised Newman-Penrose tetrad, there correspond exactly two spin-frames, differing by an overall sign only, such that
\begin{equation} \label{NPTSF}
l^a = o^A \bar{o}^{A'} \, ,~ n^a = \iota^A \bar{\iota}^{A'} \, ,~ m^a = o^A \bar{\iota}^{A'} \, ,~ \bar{m}^a = \iota^A \bar{o}^{A'} \, .
\end{equation}

We will need three \textit{normalized} Newman-Penrose tetrads that are adequate for the geometry of our spacetime $\mathcal{M}$ and its extension $\bar{\mathcal{M}}$, both of which are globally hyperbolic spacetimes. The first tetrad is defined on ${\mathcal M}$  in the coordinates $(t,x,\theta,\varphi)$ by:
\begin{equation}\label{NPTetrad}
\mathbb{T}=\begin{cases} l&= {\frac{1}{\sqrt{2f}} \left( \dl_t+ \dl_x \right)  \, ,}\\
n & = {\frac{1}{\sqrt{2f}} \left( \dl_t-\dl_x \right) \, ,}\\
m &= {\frac{1}{r\sqrt{2}} \left( \dl_{\theta} + \frac{i}{\sin \theta} \dl_{\varphi} \right) \, ,}\\
\bar{m} &= {\frac{1}{r\sqrt{2}} \left( \dl_{\theta} - \frac{i}{\sin \theta} \dl_{\varphi} \right) \, .} \end{cases}
\end{equation}
and by \eqref{NPTSF} we associate to it a spin-frame $\{ o^A , \iota^A \}$.  The tetrad $\mathbb{T}$ is adapted to the foliation $\{\Sigma_t\}_{t\in\R}$, i.e. 
\begin{equation}\label{AdaptedTetrad}
\eta_t=\frac{l+n}{\sqrt{2}}=\frac{1}{\sqrt{f}}\dl_t.
\end{equation}
This choice is taken so that the flux of the current of the Dirac field across $\Sigma_t$ defines a simple $L^2$ norm on it, as we will presently see. However, this tetrad is singular at the horizons, and a rescaling of its real vectors is needed before extending to the boundary. So, we define the tetrad $\hat{\mathbb{T}}$ on $\mathcal{M}_v$ in the chart $(v,r,\theta,\varphi)$, and the tetrad $\tilde{\mathbb{T}}$ on $\mathcal{M}_u$ in the chart $(u,r,\theta,\varphi)$, as follows:
\begin{equation}\label{NPtetradhorizon}
\hat{\mathbb{T}}=\begin{cases}
 \hat{l}&=\sqrt{f} l= \sqrt{2} \dl_v -\frac{f}{\sqrt{2}}\dl_r \, ,\\
\hat{n} & = \frac{1}{\sqrt{f}}n= {-\frac{1}{\sqrt{2}} \dl_r  \, ,}\\
\hat{m} &=m\, ,\\
\bar{\hat{m}} &=\bar{m}\, , \end{cases} %
\hspace{0.5in} 
\tilde{\mathbb{T}}=\begin{cases} 
\tilde{l}&=\frac{1}{\sqrt{f}}l=-\frac{1}{\sqrt{2}} \dl_r  \, ,\\
\tilde{n} & =\sqrt{f}n =\sqrt{2} \dl_u -\frac{f}{\sqrt{2}} \dl_r \, ,\\
\tilde{m} &=m\, ,\\
\bar{\tilde{m}} &=\bar{m}\, , \end{cases}
\end{equation} 
and we denote by $\{\hat{o}^A \, ,\hat{\iota}^A \}$ and $\{ \tilde{o}^A \, ,\tilde{\iota}^A \}$ respectively their associated spin-frames. Clearly, the tetrad $\hat{\mathbb{T}}$ (and thus its spin-frame) is regular at $\hlm$ and at $\hrp$, while the tetrad $\tilde{\mathbb{T}}$ is regular at the other two horizons' parts. 

These two spin-frames are related to the spin-frame associated to $\mathbb{T}$ by
\begin{equation}\label{hatandtildecomponents}
	\hat{o}^A = f^{1/4} o^A\, ,\hat{\iota}^A = \frac{1}{f^{1/4}} \iota^A \, ,\tilde{o}^A = \frac{1}{f^{1/4}}  o^A\, ,\tilde{\iota}^A = f^{1/4} \iota^A \, .
\end{equation}
and the components of a spinor $\phi_A$ in these  spin-frames are related to those in \eqref{componentsintetrad} by
\begin{equation}\label{relationbetweencomponents}
\hat{\phi}_0 = f^{1/4}\phi_0\, ,\hat{\phi}_1 = \frac{1}{f^{1/4}} \phi_1 \, ,\tilde{\phi}_0 = \frac{1}{f^{1/4}}  \phi_0\, ,\tilde{\phi}_1 = f^{1/4} \phi_1 \, .
\end{equation}

% Given $\cal M$ a smooth manifold, $\cal O$ an open set of $\cal M$ and $F$ a fiber bundle over $\mathcal{M}$, we denote by ${\cal C}^\infty_0 ({\cal O}\, ;~ F)$ the space of smooth sections of $F$ with compact support in ${\cal O}$.

\section{Scattering of Dirac Fields}\label{Sect:Scattering}

\subsection{The Dirac Equation and its Current}

The charged and massive Dirac equation 
%with mass $m$ and charge $qQ$. Here $Q$ is the charge of the black hole if it is charged and we allow $m=0$ and/or $q=0$. The Dirac equation 
can be written as two coupled equations on two spinors %$\tilde{\phi}_A$ and $\tilde{\chi}_{A'}$:
\begin{equation}\label{Diraceqspinor} 
\begin{cases} {\left( \nabla^{AA'} - iq A^{AA'} \right) \phi_A }  &=  {\frac{m}{\sqrt{2}} \chi^{A'} \, ,} \\ {\left( \nabla_{AA'} - i q A_{AA'} \right) \chi^{A'}} &= - {\frac{m}{\sqrt{2}} \phi_A \, .} \end{cases}
\end{equation}
Here $A^a$ is a 1-form potential of an ambient electrostatic Maxwell field, while $q$ and $m$ are respectively the charge and the mass of the Dirac field. 

Since we have already obtained a complete scattering theory for \eqref{Diraceqspinor} in  \cite{hafner_scattering_2020}, and since the purpose of the current work is to obtain conformal scattering specifically using the method of waves re-interpretation, there is no real interest in treating the scattering of a general charged massive Dirac fields. Nonetheless, the general case can be perfectly treated with the waves re-interpretation method: On the one hand, for a charged field, the same usual procedure of gauge transformation used in \cite{hafner_scattering_2020} to remove problematic behaviour arising at the horizons can be used here, and so repeating it would be redundant. On the other hand, the reader will see that adding the mass and the charge will only add first and zero order regular terms to the  second order equations (see \eqref{Omega0operator} and \eqref{Omega1operator}) relevant to our method here, without the need to change the nature of the arguments thanks to the generality of the results of \cite{hormander_remark_1990}. 

Therefore, for simplicity and for more clarity in the method, we take $q=m=0$, and generalizing to the charged massive cases is then straightforward and uses no new techniques. Accordingly, the system \eqref{Diraceqspinor} reduces to a single equation, as the second one becomes the conjugate of the first. This is the uncharged and massless Dirac equation, also known as the Weyl equation
\begin{equation}\label{Dirac-Weyleqspinor} 
 \nabla^{AA'}  \phi_A  =  0 \,.
\end{equation}

An explicit expression of \eqref{Dirac-Weyleqspinor} can be obtained using the Newman-Penrose formalism\footnote{See for instance \cite{chandrasekhar_mathematical_1984}}. For the purpose of   carrying out the specific calculations of section \ref{Section:Gourast}, we only need the expression in the tetrad $\hat{\mathbb{T}}$: 
\begin{equation}\label{NPformalism}
\begin{aligned}
\hat{n} \hat{\phi}_0 - m \hat{\phi}_1 + \hat{\mu} \hat{\phi}_0 - \hat{\beta}\hat{\phi}_1 &= 0 \, ,  \\
\hat{l} \hat{\phi}_1 - \bar{m} \hat{\phi}_0 + \hat{\alpha}\hat{\phi}_0 + (\hat{\varepsilon} -
\hat{\rho} ) \hat{\phi}_1 &= 0 \, , 
\end{aligned}
\end{equation}
where 
\begin{equation}\label{SpinCoeffs}
\hat{\rho} = \frac{f}{r\sqrt{2}}  \, ,~  \hat{\mu} =- \frac{1}{r\sqrt{2}}   \, ,~ \hat{\varepsilon} =- \frac{f'}{2\sqrt{2}} \, , \hat{\alpha} = -\hat{\beta} = -\frac{\cot \theta}{2 r \sqrt{2}} \, . 
\end{equation}
are the non-zero spin-coefficients\footnote{The spin-coefficients are the decomposition of the connection coefficients in the basis given by the Newman-Penrose tetrad. For the precise definition see \cite{hafner_scattering_2020} or directly \cite{penrose_spinors_1987}.} in this tetrad ( the rest being $\hat{\gamma}=\hat{\kappa} = \hat{\sigma} =\hat{\lambda} = \hat{\tau} = \hat{\nu} = \hat{\pi} = 0$).

The causal and future oriented\footnote{This can be seen from \eqref{componentdecompofspinor}.} vector field $J^a=\phi^A\bar{\phi}^{A'}$, referred to as the current of the Dirac field, is clearly conserved for this equation, that is
\begin{equation}\label{currentconservation}
\nabla_a J^a=0.
\end{equation}
As a matter of fact, this is also true for the general equation \eqref{Diraceqspinor} with the current $\phi^A\bar{\phi}^{A'}+\chi^A\bar{\chi}^{A'}$ which is invariant under the gauge transformation used for charged fields (see for example \cite{hafner_scattering_2020}).

The flux across a hypersurface $S$ of the current vector field $J^a=\phi^A\bar{\phi}^{A'}$ of a Dirac field $\phi_A$ defined on $\bar{\mathcal{M}}$ is
\begin{equation}\label{currentfluxS}
C(S)=\int_{S} J \hook \mathrm{d}V_{\mathbf{g}}= \int_S J_a \eta_S^a (\tau_S \hook \mathrm{d}V_{\mathbf{g}}) \; ,
\end{equation}
where $\mathrm{d}V_{\mathbf{g}}$ is the 4-volume form associated to the metric $\mathbf{g}$, while $\eta_S$ and $\tau_S$ are vector fields respectively normal and  transverse to $S$ such that $\mathbf{g}(\eta_S,\tau_S)=1$. Note that when $S$ is spacelike and future oriented, $C(s)$ is positive definite and defines an $L^2$-norm for sections of the spin-bundle over $S$. In particular, for $\Sigma_t$, we take $\eta_{\Sigma_t}=\tau_{\Sigma_t}=\eta_t$, and we can see that, in the spin-frame $\{ o^A , \iota^A \}$ of the tetrad $\mathbb{T}$, \eqref{AdaptedTetrad} and \eqref{NPTSF} imply
\begin{equation} \label{FluxSigmat}
C(t):=C(\Sigma_t)=\int_{\Sigma_t} J_a \eta_t^a (\eta_t \hook \mathrm{d}V_{\mathbf{g}}) = \frac{1}{\sqrt{2}} \int_{\Sigma_t} \left( \vert \phi_0 \vert^2 + \vert \phi_1 \vert^2 \right) \mathrm{dVol}_{\Sigma_t} \, ,
\end{equation}
where $\mathrm{dVol}_{\Sigma_t}:=(\eta_t \hook \mathrm{d}V_{\mathbf{g}})$ is the induced $3$-volume form on $\Sigma_t$. More explicitly, 
\begin{equation} \label{explicitcurrentfluxfor-t}
C(t)=\frac{1}{\sqrt{2}}\int_{\R_x\times\{t\}\times\Sphw} \left( \vert \phi_0 \vert^2 + \vert \phi_1 \vert^2 \right) r^2\sqrt{f} \d x \d \omega \, ,
\end{equation}
with $\d \omega$ the Lebesgue measure on the euclidean 2-sphere $\Sph$. We define the $L^2$-space ${\cal H}_t$ for each $t\in \R$ by
\begin{equation} \label{SpaceHt}
{\mathcal H}_t := L^2 (\Sigma_t \, ;~ \S_A  ) \, ,~ \Vert \alpha_A  \Vert^2_{{\mathcal H}_t} =\frac{1}{\sqrt{2}}\int_{\R_x\times\Sphw} \left( \vert \alpha_0 \vert^2 + \vert \alpha_1 \vert^2 \right) r^2\sqrt{f} \d x \d \omega \, ,  \, .
\end{equation}

For the horizons, the real vectors of the tetrads $\hat{\mathbb{T}}$ and $\tilde{\mathbb{T}}$ are the required normal and transverse vector fields used in \eqref{currentfluxS}. With this, we can write down the current flux across the horizons:
\begin{align}
C(\hlm) &= \int_{\hlm} J_a \hat{l}^a (\hat{n} \hook \mathrm{d}V_{\mathbf{g}}) =\frac{1}{\sqrt{2}}\int_{\R_v \times \Sphw}  \vert \hat{\phi}_0 \vert^2    r^2 \d v \d^2 \omega \, ,  \\
C(\hrm) &= \int_{\hlm} J_a \tilde{n}^a (\tilde{l} \hook \mathrm{d}V_{\mathbf{g}}) =\frac{1}{\sqrt{2}}\int_{\R_u \times \Sphw}  \vert \tilde{\phi}_1 \vert^2    r^2 \d u \d^2 \omega \, , 
\end{align}
and
\begin{equation}\label{fluxacrosshorizon}
C(\scrh_{r_-})=C(\hlm)+ C(\hrm).
\end{equation}
Similar expressions can be obtained on the past horizon. We define the following $L^2$-space on $\scrh_{r_-}$:
\begin{equation} \label{SpaceHpm}
{\mathcal H}^{+} := L^2 (\hlm)\oplus L^2 (\hrm) \, ,~ \Vert (\alpha_0,\alpha_1)  \Vert^2_{{\mathcal H}^+} =\frac{1}{\sqrt{2}}\int_{\R_v\times\Sphw}  \vert \alpha_0 \vert^2 r^2 \d v \d \omega + \frac{1}{\sqrt{2}}\int_{\R_u\times\Sphw}  \vert \alpha_1 \vert^2 r^2 \d u \d \omega
\end{equation} 
The space ${\mathcal H}^{-} $ can be define analogously on $\scrh_{r_+}$.

\subsection{Trace Operators}

%The name conformal scattering originally comes from the conformal techniques used when the scattering hypersurfaces are at infinity for the physical metric. In this case a conformal rescaling is used and the construction is done on an extended spacetime. The main objects to construct in conformal scattering are the trace operators, and consequently the scattering operator, and to establish that they are isometries between some initial and scattering data spaces on different hypersurfaces. For this, geometric energy estimates are used to show injectivity and norm preserving properties, while  showing that the operators are surjective boils down to solving the Goursat problem. In the case where no rescaling is needed, but the method still follows the rest of the above geometric construction of scattering, the     

The geometric energy estimates needed for conformal scattering are immediate from the conservation of the current \eqref{currentconservation} and the divergence theorem. In fact, since equation \eqref{Dirac-Weyleqspinor} (and \eqref{Diraceqspinor} for that matter) is a hyperbolic equation on the globally hyperbolic spacetime $\bar{\mathcal{M}}$, the standard theory for hyperbolic equations\footnote{For example, by \cite{leray_hyperbolic_1955}. This is also discussed in \cite{hafner_scattering_2020}.} together with \eqref{currentconservation} guaranty the existence of the trace operators as partial isometries between  $\mathcal{H}_0$ and $\mathcal{H}^{\pm}$, which we state in the following proposition.

\begin{prop1} \label{Prop:existenceoftraceops}
	Let ${\alpha}_A  \in \mathcal{C}^\infty_0 (\Sigma_0 \, ;~ \S_A )$, then there exist a unique ${\phi}_A  \in \mathcal{C}^\infty (\bar{\mathcal{M}}\,;~\S_A)$ solution to \eqref{Dirac-Weyleqspinor} whose restriction on $\Sigma_0$ is equal to ${\alpha}_A$. Moreover, for any $t\in\R$
	\begin{equation} \label{normpreserving}
	\Vert {\alpha}_A  \Vert_{\mathcal{H}_0}^2= C(t)= C(\scrh_{r_-})=C(\scrh_{r_+}),
	\end{equation}
	where $C(t)$ given in \eqref{FluxSigmat}, $C(\scrh_{r_-})$ and $C(\scrh_{r_+})$ given by \eqref{fluxacrosshorizon}, are the current flux of the solution $\phi_A$ across $\Sigma_t$, the future, and past horizons, respectively. Consequently, the future  trace operator $\mathbf{T}^+$ defined by
	\begin{equation}\label{futuretraceop}
	\begin{aligned}
	\mathbf{T}^+ \, :~ {\cal C}^\infty_0 (\Sigma_0 \, ;~ \S_A ) &\longrightarrow {\cal C}^\infty (\hlm)\times   {\cal C}^\infty (\hrm)\\
	 \alpha_A&\longrightarrow (\hat{\phi}_0\vert_{\hlm},\tilde{\phi}_{1}\vert_{\hrm})   
	\end{aligned}
	\end{equation} 
	and the past trace operator $\mathbf{T}^-$, defined analogously, extend to norm-preserving linear operators from $\mathcal{H}_0$ respectively to $\mathcal{H}^+$ and $\mathcal{H}^-$. Here $\hat{\phi}_0$ and $\tilde{\phi}_{1}$ are the components given in \eqref{hatandtildecomponents} of the solution $\phi_A$ associated to $\alpha_A$.
\end{prop1}

Once we prove that the trace operators are actually full isometries, i.e. surjective as well, we then have scattering: 

\begin{thm1}\label{theoremScattering}
The scattering operator given by
$
\mathbf{S}=\mathbf{T}^+(\mathbf{T}^-)^{-1}\; 
$	is an isometry from $\mathcal{H}^-$ to $\mathcal{H}^+$.
\end{thm1}

Proving that the trace operators are surjective is the aim of the next section, which is the main content of this paper. In it, we solve the Goursat problem by the waves re-interpretation method. 	
	
\section{The Goursat Problem}\label{Section:Gourast}

The general strategy for solving the Goursat problem on the inner horizons will follow the previous  work of the author  \cite{mokdad_conformal_2019} with minor modifications needed to accommodate for the different position of the singularity at $i^+$. The main steps of the proof can be summarised as follows.

Step one consists of finding the wave equations associated to the Dirac equation. This is essentially applying a Dirac operator to the Dirac operator to obtain a wave operator acting on the spinor components (see \eqref{Omega-expressions}) and leads to a system of coupled wave equations when the spinor is a Dirac spinor.  Next, the constraint equations of the Dirac system on the horizons enables us to generate from the Goursat data of the Dirac equation the Goursat data for the wave equations of step one. By embedding our spacetime -- except for a neighbourhood of $i^+$-- in a globally hyperbolic and spatially compact spacetime we obtain a solution for the wave equations using \cite{hormander_remark_1990}. We then re-interpret this solution as a Dirac field by showing that the left-hand-sides of \eqref{NPformalism} themselves satisfy a system of wave equations, related to the first one by a third application of a Dirac operator. After that we show that left-hand-sides of \eqref{NPformalism} vanish at the horizons. Well-posedness then entails that they vanish everywhere. The last step is simply extending the solution to the neighbourhood we removed before the embedding. 
\begin{figure}
	\centering
	\includegraphics[width=\textwidth]{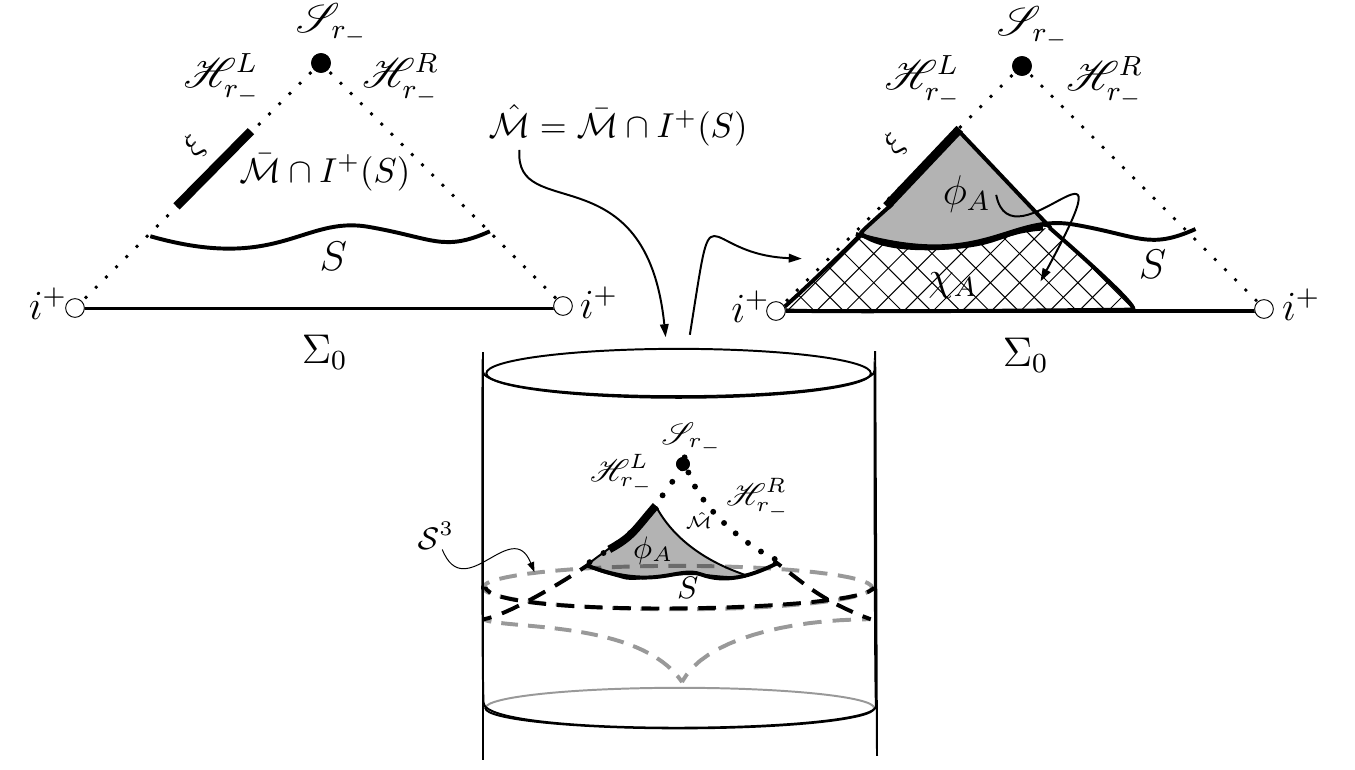}
	\caption{\emph{The resolution of the Goursat problem.}}
	\label{fig:Goursatprob}
\end{figure}  

Let us start by restating the characteristic Cauchy problem we aim to solve. We will  work out the details for the future trace operator only, the past operator being completely analogous. From Proposition \ref{Prop:existenceoftraceops}, we know that the Goursat data for the Dirac equation on the horizon are elements of $\mathcal{H}^+$, and for each element $(\xi_0,\xi_1)$ we need to find a solution for the Dirac equation on $\bar{\mathcal{M}}$ such that its restriction to $\scrh_{r_-}$ as defined in \eqref{futuretraceop}  is $(\xi_0,\xi_1)$. But by density, it is sufficient to take data in $\mathcal{H}^+$ that are smooth and compactly supported. In fact, by the linearity of the  problem, it will be enough for the data to be supported on one part of the horizon, say $\hlm$, i.e. data of the form $(\xi,0)\in\mathcal{H}^+$ where $\xi\in\mathcal{C}^\infty_0(\hlm)$.  Because the analogous case on $\hrm$ can be treat in the same way, we only give the details for $\hlm$. So, for  $\xi\in\mathcal{C}^\infty_0(\hlm)$ and using the same notations as in Proposition \ref{Prop:existenceoftraceops}, we will study the Dirac Goursat problem
\begin{equation}\label{DiracGoursatprob}
	\begin{cases}
	 \nabla^{AA'}  \phi_A  =  0 \\
	 (\hat{\phi}_0\vert_{\hlm},\tilde{\phi}_{1}\vert_{\hrm})=(\xi,0)\, .
	\end{cases}
\end{equation}

\subsection{Wave Equations} 

We rewrite \eqref{NPformalism}, the Dirac equations in the Newman-Penrose formalism, as:
\begin{equation}\label{E=0}
%\mathbf{E}(\hpz,\hpo)=
\begin{cases}
	E_0(\hpz,\hpo)=0 \, ,  \\
E_1(\hpz,\hpo)=0 \, ,
\end{cases}
\end{equation}
where
\begin{equation}\label{E-expressions}
\begin{aligned}
E_0(\hpz,\hpo):=&\hat{n} \hat{\phi}_0 - m_2 \hat{\phi}_1 - \frac{1}{r\sqrt{2}} \hat{\phi}_0 \, ,  \\
E_1(\hpz,\hpo):=&\hat{l} \hat{\phi}_1 - \bar{m}_2 \hat{\phi}_0   -\frac{1}{2r\sqrt{2}}( rf' +2f) \hat{\phi}_1  \, ,
\end{aligned}
\end{equation}
and $m_2=m +\frac{\cot \theta}{2 r \sqrt{2}}$ with $\bar{m}_2$ its conjugate. We will sometimes drop the $(\hpz,\hpo)$ in the $E_i(\hpz,\hpo)$. In addition, we set $M_2=r\sqrt{2}m_2$ and its conjugate $\bar{M}_2$, and we note that 
\begin{equation}\label{spinweightedlaplacian}
M_2\bar{M}_2=\Delta_{\Sph}^{[-\hf]}=\Delta_{\Sph} + i\frac{\cot(\theta)}{\sin(\theta)}\dl_{\varphi}-\left(\hf+\frac{1}{4}\cot(\theta)^2\right) 
\end{equation}
is the spin-weight$(-\hf)$ Laplacian on the sphere $\Sph$ in the spherical coordinates, with $\Delta_{\Sph}$ the usual Laplacian on it\footnote{Incidentally, this further emphasizes that singularities at $\theta=0,\, \pi$ are mere spherical coordinates singularities.}. We also recall the expression of the d'Alembertian of the metric $\mathbf{g}$ in the coordinates $(t,x,\theta,\varphi)$:
\begin{equation}\label{dalembertianofmetric}
\square_{\mathbf{g}}=\frac{1}{f}\left(\dl_t^2 -\dl_x^2\right)-\frac{2}{r}\dl_t-\frac{1}{r^2}\Delta_{\Sph}\, .
\end{equation}

The system of wave equations that we need, is obtained by applying essentially the same operations in $E_0$ and $E_1$, to $E_0$ and $E_1$ themselves, that is, we consider 
\begin{equation}\label{Omega-expressions}
\begin{aligned}
\Omega_0(\hpz,\hpo):=&\frac{\hat{l}}{r\sqrt{2}}( r\sqrt{2}E_0) + m_2 {E}_1 \, ,  \\
\Omega_1(\hpz,\hpo):=&\frac{\hat{n}}{r\sqrt{2}} (r\sqrt{2}E_1) + \bar{m}_2 {E}_0   \, , 
\end{aligned}
\end{equation}
where we have multiplied and divided by $r\sqrt{2}$ to slightly shorten the next calculations. To see that this indeed gives rise to a system of wave equations on $\hat{\phi}_0$ and $\hat{\phi}_1$, we calculate
\begin{align}
	\Omega_0&=\frac{\hl}{r\sqrt{2}}( r\sqrt{2}\hn\hpz)- m_2 \hl\hpo - \frac{\hl}{r\sqrt{2}} \hat{\phi}_0 +m_2\hl\hpo -m_2\bar{m}_2\hpz -\frac{1}{2r\sqrt{2}}( rf' +2f) m_2\hpo \\
	&= \hl\hn\hpz -\frac{1}{r\sqrt{2}}(\hl+f\hn)\hpz -m_2\bar{m}_2\hpz -\frac{1}{2r\sqrt{2}}( rf' +2f) m_2\hpo \\
	&= \frac{1}{2f}\left(\dl_t^2 -\dl_x^2\right)\hpz+ \frac{f'}{\sqrt{2}}\hn\hpz-\frac{1}{r}\dl_t \hpz -m_2\bar{m}_2\hpz -\frac{1}{2r\sqrt{2}}( rf' +2f) m_2\hpo \\
	&=\hf \left(\square_{\mathbf{g}}\hpz - i\frac{\cot(\theta)}{r^2\sin(\theta)}\dl_{\varphi}\hpz+\frac{1}{r^2}\left(\hf+\frac{1}{4}\cot(\theta)^2\right)\hpz + \sqrt{2}f'\hn\hpz -\frac{1}{r\sqrt{2}}( rf' +2f) m_2\hpo\right) \label{Omega0operator}
\end{align}
using  \eqref{spinweightedlaplacian} and \eqref{dalembertianofmetric} for the last equality. Similarly, we get
\begin{equation}\label{Omega1operator}
\Omega_1=\hf \left(\square_{\mathbf{g}}\hpo + i\frac{\cot(\theta)}{r^2\sin(\theta)}\dl_{\varphi}\hpo+\frac{1}{r^2}\left(\hf+\frac{1}{4}\cot(\theta)^2\right)\hpo - \frac{f'}{\sqrt{2}}\hn\hpo -\frac{1}{2r}( rf'' +3f')\hpo -\frac{\sqrt{2}}{r}\bar{m}_2\hpz\right)
\end{equation}
Thus, the wave system we will be using is
$
\mathbf{\Omega}(\hpz,\hpo)=0
$
where
$
\mathbf{\Omega}(\hpz,\hpo):=(\Omega_0(\hpz,\hpo),\Omega_1(\hpz,\hpo)) \,.
$

To construct the Goursat data for this wave system, we use the Dirac constraint equation. First we note that since $\hl$ is tangent to $\hlm$, $E_1$ is tangent to $\hlm$, and so the constraint equation on it is $E_1\vert_{\hlm}=0$, i.e.,
\begin{equation}\label{constraintequationRISTRICTIONform}
\hat{l}\vert_{\hlm} \hat{\phi}_1\vert_{\hlm} - \bar{m}_2\vert_{\hlm} \hat{\phi}_0\vert_{\hlm}   -\frac{f'(r_-)}{2\sqrt{2}} \hat{\phi}_1\vert_{\hlm}=0 \, , 
\end{equation} 
or in the $(v,r,\theta,\varphi)$ coordinates
\begin{equation}\label{constraintequationCOORDform}
\sqrt{2}\dl_v\hat{\phi}_1\vert_{\hlm}  -\frac{1}{r_-\sqrt{2}}\bar{M}_2 \hat{\phi}_0\vert_{\hlm}   -\frac{f'(r_-)}{2\sqrt{2}} \hat{\phi}_1\vert_{\hlm}=0 \, , 
\end{equation}
Let $\xi\in\mathcal{C}^\infty_0(\hlm)$ so that $(\xi,0)$ is our Dirac Goursat data. We now construct a  $\zeta\in\mathcal{C}^\infty(\hlm)$ as the solution to the constraint equation \eqref{constraintequationCOORDform} in the following (transport) initial-value problem on $\hlm$
\begin{equation}\label{constraintIVP}
\begin{cases}
\sqrt{2}\dl_v\zeta     -\frac{f'(r_-)}{2\sqrt{2}} \zeta=\frac{1}{r_-\sqrt{2}}\bar{M}_2 \xi \\
\zeta\vert_{S_p}=0
\end{cases}
\end{equation}
where $S_p$ is any 2-sphere of $\hlm$ (i.e. $\{v=cst,r=r_-\}$) in the future of the support of $\xi$. We chose $\zeta$ to be supported away from the bifurcation sphere $\scrs_{r_-}$ to avoid matching problems when we construct the Dirac solution on $\hrm$, i.e., to ensure that it will have the correct trace there. The wave Goursat problem associated to \eqref{DiracGoursatprob} is 
\begin{equation}\label{waveGoursatprob}
\begin{cases}
\mathbf{\Omega}(\hpz,\hpo) =  0 \\
(\hat{\phi}_0\vert_{\hlm},\hat{\phi}_{1}\vert_{\hlm})=(\xi,\zeta)\, ,
\end{cases}
\end{equation}
with $\zeta$ given by \eqref{constraintIVP}. 

\paragraph{Adapting to Hörmander's framework.}  In \cite{hormander_remark_1990}, Hörmander considered on a globally hyperbolic spatially compact spacetime a single scalar wave equation of the form  
\begin{equation} \label{WaveEqHormander}
\square u + L_1 u = 0
\end{equation}
with $L_1$ is a general lower order operator, and showed that the Goursat problem for this equation is well-posed. In our case, we have a system of wave equations that are coupled, but the coupling happens only on the lower order terms and not on the second order box operator, in other words, the box is only on the diagonal. Generalising Hörmander's result to such a system of wave equations is rather simple. The general operator $L_1$ of first and zero order terms in the equation \eqref{WaveEqHormander} considered in \cite{hormander_remark_1990} is controlled by a priori estimates giving exponential bounds. If the lower order term $L_1$ is a matrix instead of a simple scalar potential, it can be controlled in the same manner, and the proof goes through unchanged. This modification is minor and the required computation can be found in \cite{pham_peeling_2017}. Henceforth, when we use and refer to the results of \cite{hormander_remark_1990}, we thereby mean their generalisation to such systems. 

\begin{note}\label{Remarkhormanderresults} To get a solution for \eqref{waveGoursatprob} using Hörmander's method \cite{hormander_remark_1990}, we need to put \eqref{waveGoursatprob} into the suitable framework. Following J.-P. Nicolas \cite{nicolas_conformal_2016}, we avoid the singularities at $i^+$ by taking a spacelike hypersurface $S$ that intersects $\hrm$ in the future of $i_+$, and intersects $\hlm$ in the future of $i_+$ but in the past of the support of $\xi$. See Figure \ref{fig:Goursatprob}. 
	
Let $I^+(S)$ be the future set of $S$ and consider $\hat{\mathcal{M}}:=\bar{\mathcal{M}}\cap I^+(S)$. We now embed $(\hat{\mathcal{M}},\mathbf{g}\vert_{\hat{\mathcal{M}}})$ in a globally hyperbolic spatially compact spacetime, say a cylinder $(\R\times \mathcal{S}^3,\mathfrak{g})$. We extend $\scrh_{r_-}\cap I^+(\mathcal{S})$ as a null hypersurface that is the graph of a Lipschitz function over $\mathcal{S}^3$, and we extend $(\xi,\zeta)$ constructed above, by zero over $\hrm$ and smoothly otherwise over the extended null hypersurface. Now the extended version of  \eqref{waveGoursatprob} has a unique and smooth solution on the cylinder which we restrict to $\hat{\mathcal{M}}$ to obtain the solution of \eqref{waveGoursatprob} that we seek. 
\end{note}
In what follows, we will need the well-posedness of systems similar to \eqref{waveGoursatprob} which can be guaranteed using the same procedure we just described.

\subsection{Waves Re-interpretation}

By the discussion at the beginning of this section, we only need to solve the Goursat problem \eqref{DiracGoursatprob}. We now proceed to show that the components of the solution to the associated wave Goursat problem \eqref{waveGoursatprob} are in fact the components of a Dirac spinor in $\hat{\mathbb{T}}$, solving \eqref{DiracGoursatprob}.

\begin{thm1}\label{theorem:Goursat}
	For  $\xi\in\mathcal{C}^\infty_0(\hlm)$, there is a unique and smooth Dirac field defined on $\bar{\mathcal{M}}$, with finite current flux across $\Sigma_t$ for all $t\in\R$, and which solves \eqref{DiracGoursatprob}.
\end{thm1}
\begin{proof}
	
Finite current flux is immediate from the law of conservation of the current \eqref{normpreserving}, and thus uniqueness follows directly from the injectivity of the future trace operator and the well-posedness of the Cauchy problem on $\bar{\mathcal{M}}$.	
	
Let $\zeta$ be given by \eqref{constraintIVP} and consider the corresponding wave Goursat problem \eqref{waveGoursatprob}. By Remark \ref{Remarkhormanderresults}, there exists  $(\hpz,\hpo)$ a unique smooth solution for \eqref{waveGoursatprob}, defined on $\hat{\mathcal{M}}$. Note that from the construction in Remark \ref{Remarkhormanderresults}, $\hat{\Phi}$ is zero on $\hrm$. Actually, by finite propagation speed and local uniqueness, $(\hpz,\hpo)$ vanishes in a neighbourhood of $\hrm\cup\scrs_{r_-}$.

Now let $(E_0(\hpz,\hpo),E_1(\hpz,\hpo))$ be as in \eqref{E-expressions} and recall the decomposition \eqref{Omega-expressions} of the wave operator $\mathbf{\Omega}(\hpz,\hpo):=(\Omega_0(\hpz,\hpo),\Omega_1(\hpz,\hpo))$ of \eqref{waveGoursatprob}. Rewriting \eqref{Omega-expressions} as\footnote{Notice the similarities between \eqref{Omega-expressionsEXPANDED} and \eqref{E-expressions}. This is essentially the Dirac operator (with perturbations) applied twice to a spinor $\phi_A$ to give a wave operator on $\phi_A$. Notice also that applying a Dirac operator once more, like in \eqref{W-expressions}, gives us a wave operator on $(E_0,E_1)$, i.e., on the LHS of the Dirac equation, $ \nabla^{AA'}  \phi_A$.} 
\begin{equation}\label{Omega-expressionsEXPANDED}
\begin{aligned}
\Omega_0=&\hat{l}E_0 + m_2 {E}_1 - \frac{f}{r\sqrt{2}}E_0 \, ,  \\
\Omega_1=&{\hat{n}}E_1 + \bar{m}_2 {E}_0  -\frac{1}{r\sqrt{2}}E_1  \, , 
\end{aligned}
\end{equation}
we easily see that, similar to \eqref{Omega0operator} and \eqref{Omega1operator}, the operator $\mathbf{W}(E_0,E_1):=(W_0(E_0,E_1),W_1(E_0,E_1))$  given by\footnote{We could have multiplied and divided by $r\sqrt{2}$ to get expressions that are more similar to \eqref{Omega-expressionsEXPANDED}, but this does not change the argument.}  
\begin{equation}\label{W-expressions}
\begin{aligned}
W_0:=&{\hat{n}}\Omega_0 - {m}_2 \Omega_1 \, ,  \\
W_1:=&\hat{l}\Omega_1 - \bar{m}_2 \Omega_0  \, , 
\end{aligned}
\end{equation}	
is a wave operator on $(E_0,E_1)$. Since $(\hpz,\hpo)$ is a solution of \eqref{waveGoursatprob}, we have $\mathbf{\Omega}(\hpz,\hpo)=0$, which in turn implies that $\mathbf{W}(E_0,E_1)=0$. If in addition we show that $(E_0,E_1)$ vanishes on $\scrh_{r_-}$, then it solves the wave Goursat problem 
\begin{equation}\label{TheW-waveGoursatprobonE0E1}
\begin{cases}
\mathbf{W}(E_0,E_1) =  0 \, , \\
(E_0,E_1)\vert_{\scrh_{r_-}}=0\, ,
\end{cases}
\end{equation}
and by uniqueness (in \cite{hormander_remark_1990}) we must have $(E_0,E_1)=0$ on $\hat{\mathcal{M}}$. In other words, we get \eqref{E=0} and hence $(\hpz,\hpo)$ will indeed be the components of a Dirac field which, as we will presently show, extends to the rest of the future of $\Sigma_0$ and solves \eqref{DiracGoursatprob}, as required. 

First, we note that since $(\hpz,\hpo)$ is zero on $\hrm$, we already have $(E_0,E_1)\vert_{\hrm}=0$. Also, we know that $E_1\vert_{\hlm}=0$ is the constraint equation on $\hlm$, which also entails in particular that ${m}_2\vert_{\hlm} E_1\vert_{\hlm}=0$ since $m_2$ is tangent to $\hlm$. It remains to show that $E_0\vert_{\hlm}=0$. From \eqref{waveGoursatprob}, we have
\begin{equation}
\Omega_0=\hat{l}E_0 + m_2 {E}_1 -\frac{f}{r\sqrt{2}}E_0=0,
\end{equation}
and restricted to $\hlm$, it becomes (recall that $f(r_-)=0$)
\begin{equation}
\Omega_0\vert_{\hlm}=\hat{l}\vert_{\hlm} E_0\vert_{\hlm}=0.
\end{equation}
Finally, as $(\hpz,\hpo)$ vanishes in a neighbourhood of $\scrs_{r_-}$, there exist a 2-sphere $S_p$ of $\hlm$ (i.e. $\{v=cst,r=r_-\}$) in this neighbourhood such that $E_0\vert_{S_p}=0$. Therefore, $E_0\vert_{\hlm}$ solves a simple transport initial-value problem similar to \eqref{constraintIVP}, namely
\begin{equation}
\begin{cases}
\dl_v E_0\vert_{\hlm}=0 \\
(E_0\vert_{\hlm})\vert_{S_p}=0,
\end{cases}
\end{equation}
and hence $E_0\vert_{\hlm}=0$. 

Let $\phi_A$ denote the Dirac solution we just found and note that it does indeed solve \eqref{DiracGoursatprob} on $\hat{\mathcal{M}}$ since $\tilde{\phi}_1$, its second component in the tetrad $\tilde{\mathbb{T}}$, is equal to $f^\hf\hpo$ by \eqref{relationbetweencomponents}, and thus,  $\tilde{\phi}_{1}\vert_{\hrm}=0$.

The last step is extending the solution down to $\Sigma_0$. This is again following \cite{nicolas_conformal_2016}. First, we note that $\phi_A$ as a Dirac spinor with finite current flux, verifies $\phi_A\vert_S\in L^2\left(S; \mathbb{S}_A \right)$, and its trace on $\hlm$, $\xi$ , vanishes on $\scrh_{r_-}\cap S$ because of our choice of the hypersurface $S$ in Remark \ref{Remarkhormanderresults}. This means that we can approximate $\phi_A\vert_S$ by a sequence $\{{\psi_A^n}\}_{n\in\mathbb{N}}$ in $C_0^\infty
\left(S; \mathbb{S}_A \right)$,  i.e., supported away from the horizon. Now let $\phi^n_A$ be the smooth Dirac solution corresponding to the initial data ${\psi_A^n}$ on $S$. Since the current of $\phi^n_A$ is conserved between $S$ and $\Sigma_t$ for all $t$, the sequence $\{\phi^n_A\vert_{\Sigma_t}\}_{n\in\mathbb{N}}$ is Cauchy in $L^2\left(\Sigma_t; \mathbb{S}_A \right)$ and hence convergent. Let $\chi_A(t)$ be the limit of the sequence $\{\phi^n_A\vert_{\Sigma_t}\}_{n\in\mathbb{N}}$. By local uniqueness, $\chi_A$ and $\phi_A$ coincide in the future of $S$, therefore extending $\phi_A$ down to $\Sigma_0$. 
\end{proof}

\printbibliography[heading=bibintoc] 
\end{document}